\documentclass[aps,prd,onecolumn,groupedaddress,showpacs,nofootinbib,amssymb]{revtex4}

\usepackage[dvips]{graphicx}
\usepackage{amssymb}
\usepackage{amsmath}
\usepackage{graphicx,,color}
\usepackage{amsfonts}
\usepackage{bm}
\usepackage{epstopdf}
\usepackage{cancel}
\usepackage{comment}

\newcommand\be{\begin{equation}}
\newcommand\ee{\end{equation}}

\allowdisplaybreaks[4]

\begin{document}
\tolerance=5000

\title{Logarithmic corrected Einstein-Gauss-Bonnet inflation compatible with GW170817}
\author{  S.A.
Venikoudis,  \thanks{venikoudis@gmail.com} F.P.
Fronimos\,\thanks{fotisfronimos@gmail.com}}
\affiliation{
 Department of Physics,
 Aristotle University of
Thessaloniki, Thessaloniki 54124,
Greece\\}

\tolerance=5000

\begin{abstract}
In this paper we investigate the inflationary phenomenology of an Einstein-Gauss-Bonnet theory with the extension of a logarithmic modified $f(R)$ gravity, compatible with the GW170817 event. The  main idea of our work is to study different results for an almost linear  Ricci scalar through logarithmic corrections and examine  whether  such  model  is  viable. First of all, the theoretical framework under slow-roll evolution of the scalar field is presented and also developed the formalism of the constant-roll evolution making predictions for the non-
Gaussianities of the models is developed , since the constant-roll evolution is known to enhance non-Gaussianities.
As shown, the non-Gaussianities are of the order $\mathcal{O}\sim(10^{-1})$. Furthermore, the slow-roll indices and the observational indices of inflation, are calculated for several models of interest. As demonstrated, the phenomenological viability of the models at hand is achieved for a wide range
of the free parameters and the logarithmic term has a minor contribution to numerical calculations, as expected.
\end{abstract}


\maketitle

\section{Introduction}

The recent years, one of the most remarkable achievements in theoretical and observational Cosmology is, without a doubt, the detection of the gravitational waves about one hundred years after Einstein's predictions. 
In 2017, the LIGO-Virgo detectors observed a gravitational wave coming from the merging of two neutron stars \cite{GBM:2017lvd}. The interesting fact of the observation was that the  gravitational wave arrived almost equal with the gamma-ray burst. This means that, the speed of the gravitational wave $c_{T}$ is approximately equal to the speed of light, namely $c_{T}^2=1$ in Natural Units. This constraint imposes stringent conditions on modified theories of gravity that may describe successfully the nature on such scales. Many generalized theories of gravity provide viable models compatible with this astrophysical event, see Ref.\cite{Ezquiaga:2017ekz}.

One attractive class of theories, in order to approach the inflationary era of the universe or other astrophysical phenomena is the Einstein-Gauss-Bonnet theories \cite{Hwang:2005hb,Nojiri:2006je,Cognola:2006sp,Nojiri:2005vv,Nojiri:2005jg,Satoh:2007gn,Bamba:2014zoa,Yi:2018gse,Guo:2009uk,Guo:2010jr,Jiang:2013gza,Kanti:2015pda,Nojiri:2017ncd,Kanti:1998jd,Pozdeeva:2020apf,Fomin:2020hfh,DeLaurentis:2015fea,Chervon:2019sey,Nozari:2017rta,Odintsov:2018zhw,Kawai:1998ab,Yi:2018dhl,vandeBruck:2016xvt,Kleihaus:2019rbg,Bakopoulos:2019tvc,Maeda:2011zn,Bakopoulos:2020dfg,Ai:2020peo,Odintsov:2019clh,Oikonomou:2020oil,Odintsov:2020xji,Oikonomou:2020sij,Odintsov:2020zkl,Odintsov:2020sqy,Easther:1996yd,Antoniadis:1993jc,Antoniadis:1990uu,Kanti:1995vq,Kanti:1997br,Bajardi:2019zzs,Capozziello:2019wfi,deMartino:2020yhq,Nojiri:2010wj}. These theories involve the Gauss-Bonnet term, more specifically, quadratic terms of the Ricci scalar and of the Riemann and Ricci tensors, in the context of Einstein's gravity. Our motivation to utilize this class of gravitational theories in order to describe the inflationary era of the universe originate from the string property. In detail, the whole theory is string-corrected canonical scalar field theory minimally coupled to gravity. In our previous work \cite{Odintsov:2020sqy}, we proved that this theory can be rectified in view
of the GW170817 event, by simply setting the gravitational speed wave equal to unity and expressing all the physical quantities in terms of the scalar field.

In this paper we shall extent our previous works \cite{Odintsov:2020sqy, Odintsov:2020mkz}, presenting a modified theory of gravity $f(R)$  with a logarithmic term, based on Ref.\cite{Odintsov:2017hbk}, also modified gravity with logarithmic terms is presented in  Ref.\cite{Nojiri:2003ni}. First of all, we present in detail the theoretical framework
of the evolution of the scalar field under slow-roll and constant-roll approximations in the background of Friedmann-Roberton-Walker spacetime. As mentioned before, we express all the involved physical quantities in terms of functions
of the scalar field and their higher derivatives with respect to the scalar field. By imposing the slow-roll conditions into the gravitational equations of motion, the slow-roll indices and the observational indices have quite simple and elegant final expressions. Afterwards, we consider the constant roll evolution of the scalar field, by imposing the condition $\ddot \phi=\beta H \dot \phi$ and we repeat the same process. Moreover, the amount of Non-Gaussianities is predicted and as expected is not significantly altered in comparison to the $R$ gravity. Finally, we shall examine the compatibility of the theoretical framework with the latest Plack Data see, Ref \cite{Akrami:2018odb} in the context of the two possible ways of the scalar field evolution. In the last section, viable models for the inflationary era are constructed by considering coupling scalar field functions $\xi (\phi)$ with compatible numerical values with the observations.

Before we begin our analysis, it is necessary to explain the reason why the theory needs massless gravitons in order to be consistent with the GW170817 event. In nature the interactions between particles and fields are achieved through the propagators of the fundamental forces. The graviton is the progagator of the gravitational waves produced either in the early universe or from astrophysical events. From the perspective of the Elementary Particle Physics the graviton must be massless during the inflationary and post-inflationary era. Despite the appealing string property of the Einstein-Gauss-Bonnet theory there is a serious drawback. The theory predicts that the primordial tensor perturbations are incopatible with the GW170817 event. As we have proved in our previous works \cite{Oikonomou:2020oil} it is possible to get a massless graviton under certain constraints. Only if the coupling scalar function $\xi (\phi)$ satisfies the differential equation $\ddot \xi- H \dot \xi=0$, we can demand $c_{T}^2=1$. Hence, we can obtain a viable model for the inflationary era in the context of Einstein-Gauss-Bonnet gravity.

\section{THEORETICAL FRAMEWORK OF LOGARITHMIC CORRECTED $f(R)$ GRAVITY}
We begin our analysis by considering the following gravitational theory of a scalar field $\phi$, since all the information about the universe
in the era of inflation is encoded in it. Let us assume that the action is defined as,
\begin{equation}
\centering
\label{action}
S=\int {d^4x\sqrt{-g}\left( \frac{f(R)}{2\kappa^2}-\frac{1}{2} g^{\mu\nu}\partial_\mu\phi\partial_\nu\phi-V(\phi)-\xi(\phi)\mathcal{G}\right)},
\end{equation}
where, $g$ is the determinant of the metric tensor, $\kappa=\frac{1}{M_P}$ is the gravitational constant while $M_P$ denotes the reduced Planck mass, $V(\phi)$ is the scalar potential and $\xi(\phi)$
signifies the Gauss-Bonnet coupling scalar function. We assume a modified theory of gravity $f(R)$ where, $f(R)=R^{1+\delta}$ with $\delta\ll1$. Considering a power-law model, becomes   abundantly  clear  that  for  small  deviations  from  unity,  one  could  easily  obtain  logarithmic corrections to gravity via Taylor expansion, such terms are known for describing quantum corrections.   Finally,  the Gauss-Bonnet term $\mathcal{G}$ is given by the expression $\mathcal{G}=R^2-4R_{\alpha\beta}R^{\alpha\beta}+R_{\alpha\beta\gamma\delta}R^{\alpha\beta\gamma\delta}$,
with $R_{\alpha\beta}$ and $R_{\alpha\beta\gamma\delta}$ being the
Ricci and Riemann tensor respectively. Furthermore, the line-element is assumed to have the
Friedmann-Robertson-Walker form,
\begin{equation}
\centering
\label{metric}
ds^2=-dt^2+a(t)^2\delta_{ij}dx^idx^j,\,
\end{equation}
where a(t) is the scale factor of the Universe and the metric tensor has the form of $g_{\mu\nu}=diag(-1, a(t)^2, a(t)^2, a(t)^2)$.
The effective Lagrangian of inflation is not specified by the data
at present time. Thus, although the inflationary era is a classical
era of our Universe, which is described by a four dimensional
spacetime, it still is possible that the quantum era may have a direct
imprint on the effective Lagrangian of inflation. Therefore, the two
most simple corrections of the inflationary effective Lagrangian may
be provided by higher curvature terms , like $f(R)$ gravity
corrections, and Einstein-Gauss-Bonnet corrections.
As long as the metric is flat, the Ricci scalar and the Gauss-Bonnet term are topological invariant and can be written as $R=12H^2+6\dot H$,
$\mathcal{G}=24H^2(\dot H+H^2)$ respectively. $H$ is Hubble's parameter and in addition, the ``dot'' denotes differentiation with respect to the cosmic time. We expand the modified gravitational function $f(R)$ as follows,
\begin{equation}
\centering
\label{f(R)}
f(R)\simeq R+\delta Rln(\alpha R),
\end{equation}
where $\alpha$ is a constant with mass dimensions $[m]^{-2}$ for consistency. It is expected that the logarithmic term $ln(\alpha R)$ has minor contribution to the equations of motion because represents quantum corrections. Differentiating Eq. (\ref{f(R)}) with respect to the Ricci scalar gives,
\begin{equation}
\centering
\label{F}
F\simeq1+\delta+\delta \ln(\alpha R).
\end{equation}
Implementing the variation principle with respect to the metric tensor and the scalar field in Eq. (\ref{action}) generates the
field equations of gravity and the continuity equation of the scalar field. By splitting the field equations in time and
space components, the gravitational equations of motion are then derived which read,
\begin{equation}
\centering
\label{motion1}
\frac{3FH^2}{\kappa^2}=\frac{1}{2}\dot\phi^2+V+24\dot\xi H^3+\frac{FR-f}{2\kappa^2}-\frac{3H\dot F}{\kappa^2},\,
\end{equation}
\begin{equation}
\centering
\label{motion2}
\frac{-2F\dot H}{\kappa^2}=\dot\phi^2-16\dot\xi H\dot H-8H^2(\ddot\xi-H\dot\xi)+\frac{\ddot F-H\dot F}{\kappa^2},\,
\end{equation}
\begin{equation}
\centering
\label{motion3}
\ddot\phi+3H\dot\phi+V'+\xi'\mathcal{G}=0.\,
\end{equation}

As proved in a recent work of ours \cite{Odintsov:2020sqy}, certain additional constraints on the gravitational wave speed need to be imposed so as to achieve compatibility with recent striking observations from GW170817. Gravitational waves are perturbations in the metric which travel through spacetime with the speed of light. The gravitational wave speed in
natural units for Einstein-Gauss-Bonnet theories has the form,
\begin{equation}
\centering
c_T^2=1-\frac{Q_f}{2Q_t},\,
\end{equation}
 where $Q_f= 16(\ddot\xi-H\dot\xi)$ and $Q_t=\frac{F}{\kappa^2}-8\dot\xi H$, are auxiliary functions depending on the scalar field and the Ricci scalar.
Compatibility can be achieved by equating
the velocity of gravitational waves with unity, or making it infinitesimally close to unity. In other words, we demand $Q_f=0$. The constraint leads to an ordinary differential equation $\ddot \xi=H\dot \xi$. Now we will solve this equation in terms of the derivatives of scalar field. Assuming that $\dot \xi=\xi' \dot \phi$ and $\frac{d}{dt}=\dot \phi \frac{d}{d\phi}$ the constraint equation has the form,
\begin{equation}
\label{constraint}
\centering
\xi'' \dot \phi^2+\xi' \ddot \phi=H\xi' \dot \phi.
\end{equation}
Considering the approximation
\begin{equation}
\centering
\xi' \ddot \phi\ll\xi'' \dot \phi^2,
\end{equation}
Eq. (\ref{constraint}) can be solved easily with respect to the derivative of the scalar field,
\begin{equation}
\label{fdot}
\centering
\dot \phi \simeq \frac{H\xi'}{\xi''}.
\end{equation}
In order to study the inflationary era of the Universe it is necessary to solve analytically the system of equations of motion. It is obvious that, this system is very difficult to study analytically. Thus, we assume the slow-roll approximations during inflation. Mathematically speaking, the following conditions are assumed to hold true,
\begin{align}
\label{approx}
\centering
\dot H&\ll H^2,& \frac{1}{2}\dot\phi^2&\ll V,& \ddot\phi\ll3 H\dot\phi,\
\end{align}
thus, the equations of motion can be simplified greatly. 
Hence, after imposing the constraint of the gravitational wave and considering the slow-roll approximations the equations of motion have the following elegant forms,
\begin{equation}
\centering
\label{motion1}
\frac{3FH^2}{\kappa^2}=V+24\dot\xi H^3+\frac{FR-f}{2\kappa^2}-\frac{3H\dot F}{\kappa^2},\,
\end{equation}
\begin{equation}
\centering
\label{motion2}
\frac{-2F\dot H}{\kappa^2}=-16\dot\xi H\dot H+\frac{\ddot F-H\dot F}{\kappa^2},\,
\end{equation}
\begin{equation}
\centering
\label{motion3}
3H\dot\phi+V'+\xi'\mathcal{G}=0.\,
\end{equation}
However, even with the slow-roll approximations holding true, the system of differential equations still remains intricate
and cannot be solved. Further approximations are needed in order to derive the inflationary phenomenology, so  we neglect string corrections themselves. This is a reasonable assumption since even though the Gauss-Bonnet scalar coupling function is seemingly neglected, it participates indirectly from the gravitational wave condition. Also, in many cases, string corrections are proven to be subleading. Moreover, under slow-roll assumptions the Ricci scalar is written as, $R\simeq 12 H^2$. Recalling that $f(R)\simeq R+\delta Rln(\alpha R)$, one obtains elegant simplifications and functional expressions for the equations of motion. The first and the second derivatives of the function F are respectively, $\dot F=2\delta(\frac{\dot H}{H})$, $\ddot F=2\delta(\frac{\ddot H}{H}-\frac{\dot H^2}{H^2})$. The last two terms in the first equation of motion are quite smaller in order of magnitude than the scalar potential of the field hence, as we will prove in the fourth section numerically, these terms can be neglected. The same approximation can be applied in the second equation of motion for the last term of the right hand side. Thus, the final simplified equations of motion are,
\begin{equation}
\centering
\label{motion1final}
H^2 \simeq \frac{\kappa^2V}{3(1-\delta)},\,
\end{equation}
\begin{equation}
\centering
\label{motion2final}
\dot H \simeq -\frac{1}{2}(\kappa \dot \phi)^2,\,
\end{equation}
\begin{equation}
\centering
\label{motion3final}
3H\dot\phi+V'\simeq0,\,
\end{equation}
where the parameter $\delta$ can be discarded for $\lvert \delta \rvert\ll1$. In the next section we shall prove that the discarded terms are quite smaller in order of magnitude.
The dynamics of inflation can be described by six parameters named
the slow-roll indices, defined as follows
\cite{Hwang:2005hb,Odintsov:2020sqy},
\begin{align}
\centering \epsilon_1&=-\frac{\dot
H}{H^2},&\epsilon_2&=\frac{\ddot\phi}{H\dot\phi},&\epsilon_3&=\frac{\dot
F}{2HF},&\epsilon_4&=\frac{\dot E}{2HE},&\epsilon_5&=\frac{\dot
F+\kappa^2Q_a}{2H\kappa^2Q_t},&\epsilon_6&=\frac{\dot Q_t}{2HQ_t},\,
\end{align}
where the auxiliary functions are defined as,
\begin{align}
\centering Q_a&=-8\dot\xi H^2,&Q_b&=-16\dot\xi H,&Q_e&=-32\dot\xi\dot H,&E&=\frac{F}{\kappa^2\dot \phi^2}\left(\dot \phi^2+3\frac{(\dot F +\kappa^2Q_a)^2}{2\kappa^4Q_t}\right),\,
\end{align}
in detail, the auxiliary functions and the slow-roll indices can be written as follows,
\begin{equation}
\centering
Q_a=\frac{8\kappa^2V(\frac{\kappa^2V}{3-3\delta})^\frac{1}{2}\xi'^2}{3(\delta-1)\xi''},\,
\end{equation}
\begin{equation}
\centering
Q_b=-\frac{16\kappa^2V(\xi')^2}{3(1-\delta)\xi''},\,
\end{equation}
\begin{equation}
\centering
Q_e=- \frac{16\kappa^4V(\frac{\kappa^2V}{3-3\delta})^\frac{1}{2}(\xi')^4}{3(\delta-1)(\xi'')^3}\,
\end{equation}
\begin{equation}
\centering
Q_t=\frac{\delta +1+\delta \ln (-\frac{4 a \kappa^2 V}{\delta -1})}{\kappa^2}+\frac{8 \kappa^2 V(\xi')^2}{3 (\delta -1) \xi ''}\,
\end{equation}
\begin{equation}
\centering
\label{index1sg}
\epsilon_1\simeq\frac{\kappa^2}{2}\left(\frac{\xi'}{\xi''}\right)^2,\,
\end{equation}
\begin{equation}
\label{index2sg}
\epsilon_2\simeq 1+\frac{\xi'(V'\xi''-2V\xi''')}{2V(\xi'')^2},\,
\end{equation}
\begin{equation}
\label{index3sg}
\centering
\epsilon_3=-\frac{\kappa^2\delta(\xi')^2}{2(1+\delta+\delta  \ln \left(-\frac{4 a \kappa^2 V}{\delta -1}\right))(\xi'')^2},\,
\end{equation}
\begin{equation}
\label{index5sg}
\centering
\epsilon_5=\frac{(\kappa\xi')^2(-3(\delta-1)\delta+8V\xi''\kappa^2)}{2\xi''(8\kappa^4V(\xi')^2+3(\delta-1)(1+\delta+\delta\ln(-\frac{4 a \kappa^2 V}{\delta -1})\xi''))},\,
\end{equation}
\begin{equation}
\label{index6sg}
\centering
\epsilon_6=\frac{\xi ' \left(\kappa^2 V \xi ' \xi '' \left(8 k^2 \xi ' V'-3 (\delta -1) \delta \right)-8 \kappa^4 V^2 \xi ' \left(\xi ^{(3)} \xi '-2 (\xi '')^2\right)+3 (\delta -1) \delta  (\xi '')^2 V' \right)}{2 V (\xi '')^2 \left(3 (\delta -1) \xi '' \left(\delta  \ln \left(-\frac{4 a \kappa^2 V}{\delta -1}\right)+\delta +1\right)+8 \kappa^4 V (\xi ')^2\right)},\,
\end{equation}
where, the index $\epsilon_4$ is omitted due to the perplexed expression. The scalar potential with an unspecified scalar coupling function $\xi(\phi)$ can be written as,
\begin{equation}
\centering
\label{potential}
V(\phi)=V_0 e^{\int\frac{\kappa^2 \xi'(\phi)}{(\delta-1)\xi''(\phi)}   d\phi }.
\end{equation}
with $V_0$ being the amplitude of the scalar potential with mass dimensions $[m]^4$. In order to examine the validity of a model, the results which the model produces must be confronted to the recent
Planck observational data [47]. In the following model, we shall derive the values for the quantities, namely the
spectral index of primordial curvature perturbations $n_\mathcal{S}$, the tensor-to-scalar-ratio r and finally, the tensor spectral
index $n_\mathcal{S}$ [3, 36]. These quantities are connected with the slow-roll indices introduced previously, as shown below,
\begin{align}
\label{observed}
\centering
n_s&=1-2\frac{2\epsilon_1+\epsilon_2-\epsilon_3+\epsilon_4}{1-\epsilon_1},&n_T&=-2\frac{\epsilon_1+\epsilon_6}{1-\epsilon_1},&r&=16\left|\left(\frac{\kappa^2Q_e}{4HF^2}-\epsilon_1-\epsilon_3\right)\frac{Fc_A^3}{\kappa^2Q_t}\right|,\,
\end{align}
where  $c_A$ the sound wave velocity defined as,
\begin{equation}
\centering
\label{soundwave}
c_A^2=1+\frac{\kappa^4(\dot F+\kappa^2Q_a)Q_e}{2\kappa^4Q_t\dot \phi^2+3(\dot F+Q_a)^2}.\,
\end{equation}
Based on the latest Planck observational data \cite{{Akrami:2018odb}} the spectral index of primordial curvature  perturbations is $n_\mathcal{S}=0.9649\pm 0.0042 $
and the tensor-to-scalar-ratio r must be $r<0.064$.
Our goal now is to evaluate the observational
indices during the first horizon crossing. However, instead of
using wavenumbers, we shall use the values of the scalar potential
during the initial stage of inflation. Taking it as an input, we
can obtain the actual values of the observational quantities. We
can do so by firstly evaluating the final value of the scalar
field. This value can be derived by equating slow-roll index
$\epsilon_1$ in equation (\ref{index1sg}) to unity. Consequently, the
initial value can be evaluated from the $e$-foldings number,
defined as
$N=\int_{t_i}^{t_f}{Hdt}=\int_{\phi_i}^{\phi_f}{\frac{H}{\dot\phi}d\phi}$,
where the difference $t_f-t_i$ signifies the duration of the
inflationary era. Recalling the definition of $\dot\phi$ in Eq.
(\ref{fdot}), one finds that the proper relation from which the
initial value of the scalar field can be derived is,
\begin{equation}
\centering
\label{efolds1}
N=\int_{\phi_i}^{\phi_f}{\frac{\xi''}{\xi'}d\phi}.\,
\end{equation}
From this equation, as well as equation (\ref{index1sg}), it is
obvious that choosing an appropriate coupling function, is the key
in order to simplify the results.

\section{CONSTANT-ROLL EVOLUTION OF THE SCALAR FIELD IN LOGARITHMIC CORRECTED $f(R)$ GRAVITY AND PRIMORDIAL NON-GAUSSIANITIES}

In the following section we shall analyse the theoretical framework of the evolution of the scalar field under constant-roll condition $\ddot \phi=\beta H \dot \phi$, where $\beta$ is defined as the constant-roll parameter to be specified later. After the specific assumption and neglecting string corrections the equations of motion modified as follows,
\begin{equation}
\label{constantfdot}
\dot \phi=H(1-\beta)\frac{\xi'}{\xi''},
\end{equation}
as a result, the equations of motion are,
\begin{equation}
\centering
\label{motion1gc}
H^2 \simeq \frac{\kappa^2V}{3(1-\delta)},\,
\end{equation}
\begin{equation}
\centering
\label{motion2gc}
\dot H \simeq -\frac{1}{2}\kappa^2 H^2(1-\beta)^2(\frac{\xi'}{\xi''})^2  ,\,
\end{equation}
\begin{equation}
\centering
\label{motion3gc}
(3+\beta)H\dot\phi+V'\simeq0.\,
\end{equation}
In the context of the constant-roll evolution of the scalar field, the slow-roll indices for an arbitrary coupling scalar  function $\xi(\phi)$ are given by the following expressions,

\begin{equation}
\centering
\label{index1cg}
\epsilon_1=\frac{(\beta-1)^2 (\kappa \xi ')^2}{2 (\xi '')^2} ,\,
\end{equation}
\begin{equation}
\label{index2cg}
\epsilon_2=\beta ,\,
\end{equation}
\begin{equation}
\label{index3cg}
\centering
\epsilon_3= -\frac{(\beta-1)^2 \delta  (\kappa\xi ')^2}{2 (\xi '')^2 \left(\delta  \ln \left(-\frac{4 a \kappa^2 V}{\delta -1}\right)+\delta +1\right)},\,
\end{equation}
\begin{equation}
\label{index5cg}
\centering
\epsilon_5=-\frac{(\beta-1) \kappa^2 (\xi ')^2 \left(3 (\beta-1) (\delta -1) \delta +8 \kappa^2 V (\xi '')\right)}{2 \xi '' \left(3 (\delta -1) \xi '' \left(\delta  \ln\left(-\frac{4 a \kappa^2 V}{\delta -1}\right)+\delta +1\right)-8 (\beta-1) \kappa^4 V (\xi ' )^2\right)}, \,
\end{equation}
\begin{equation}
\label{index6cg}
\centering
\epsilon_6= \frac{(\beta-1) \xi ' \left((\beta-1)\kappa^2 V \xi ' \xi '' \left(3 (\delta -1) \delta -8 \kappa^2 \xi ' V'\right)+8 (\beta-1) \kappa^4 V^2 \xi ' \left(\xi ^{(3)} \xi '-2 (\xi '')^2\right)+3 (\delta -1) \delta  (\xi '')^2 V'\right)}{2 V( \xi '')^2 \left(8 (\beta-1) \kappa^4 V (\xi ')^2-3 (\delta -1) \xi '' \left(\delta  \ln \left(-\frac{4 a \kappa^2 V}{\delta -1}\right)+\delta +1\right)\right)},\,
\end{equation}
where again as before, the index $\epsilon_4$ is omitted due to the lengthy expression. The auxiliary functions is obvious that can be found easily. The e-folding number can be written as, 
\begin{equation}
\centering
\label{efolds}
N=\frac{1}{1-\beta}\int_{\phi_i}^{\phi_f}{\frac{\xi''}{\xi'}d\phi}.\,
\end{equation}   

When the scalar field evolves with a constant-rate of roll enhances the non-Gaussianities features. Until now, the perturbations in the Cosmic Microwave Background
(CMB) are described perfectly as Gaussian distributions, since no
practical evidence is found pointing out a non-Gaussian pattern in
the CMB. It is possible though, not evident for the moment, that
in the following years the observations may reveal a non-Gaussian
pattern in the CMB primordial power spectrum. In this section we
shall discuss how to evaluate the non-Gaussianities quantitatively
in the context of the GW170817-compatible Einstein-Gauss-Bonnet
gravity, using the formalism and notation of
\cite{DeFelice:2011zh}. Even though a $f(R)$ logarithmic gravity is assumed, we consider the same equations as the \cite{DeFelice:2011zh}. We first define the following quantities,
\begin{align}
\centering
\delta_\xi&=\kappa^2H\dot\xi,&\delta_X&=\frac{\kappa^2\dot\phi^2}{H^2},&\epsilon_s&=\epsilon_1-4\delta_\xi,&n_s&=\frac{\dot\epsilon_s}{H\epsilon_s},&s&=\frac{\dot
c_A}{Hc_A}.\,
\end{align}
Here, we shall implement a different formula for the sound wave
speed, however equivalent to the previous, which is based on these
newly defined quantities for convenience and reads,
\begin{equation}
\centering
c_A^2\simeq1-\frac{64\delta_\xi^2(6\delta_\xi+\delta_X)}{\delta_X}.\,
\end{equation}
Recalling equations (\ref{constantfdot}), (\ref{motion2gc}) and
(\ref{motion3gc}), one finds that the aforementioned auxiliary terms have the following forms for an unspecified scalar coupling function $\xi(\phi)$,
\begin{equation}
\centering
\delta_\xi\simeq \frac{(\beta-1) \kappa^4 V (\xi ')^2}{3 (\delta -1) \xi ''}\,
\end{equation}
\begin{equation}
\centering
\delta_X\simeq \frac{(\beta-1)^2  (\kappa\xi ')^2}{(\xi '')^2} ,\,
\end{equation}
\begin{equation}
\centering
\epsilon_s\simeq \frac{(\beta-1) (\xi ')^2 \left(3 (\beta-1) (\delta -1)\kappa^2-8 \kappa^4 V \xi '' \right)}{6 (\delta -1) (\xi '')^2},\,
\end{equation}
\begin{equation}
\centering
n_s\simeq \frac{2 (\beta-1) \left(\xi ^{(3)} \xi ' \left(4\kappa^2 V \xi ''-3 (\beta-1) (\delta -1)\right)-(\xi '')^2 \left(-3 (\beta-1) (\delta -1)+4 \kappa^2 \xi ' V'+8 \kappa^2 V \xi '' \right)\right)}{(\xi '')^2 \left(8 \kappa^2 V \xi ''-3 (\beta-1) (\delta -1)\right)} ,\,
\end{equation}
\begin{equation}
\centering
s=(1-\beta)\frac{\xi'}{\xi''}\frac{c_A'}{c_A}.\,
\end{equation}

In certain examples, we shall demonstrate that by choosing appropriately
the coupling function, the quantity
$\epsilon_s$ is simplified greatly, and it shall also coincide with $\epsilon_1$, along
with $\delta_X$. No matter the form of the sound wave velocity,
the derivative $c_A$ is very complex, hence for obvious reasons its analytic
expression is omitted. These forms are very useful due to the fact that the
power spectra $\mathcal{P}_S$ of the primordial curvature
perturbations and the equilateral momentum approximation term
$f_{NL}^{eq}$ can be derived from such terms. These quantities are
defined as,
\begin{equation}
\centering
\label{spectra}
\mathcal{P}_S=\frac{\kappa^4V}{24\pi^2\epsilon_sc_A},\,
\end{equation}
\begin{equation}
\centering
\label{NL}
f_{NL}^{eq}\simeq\frac{55}{36}\epsilon_s+\frac{5}{12}n+\frac{10}{3}\delta_\xi.\,
\end{equation}
In the following we shall appropriately specify the value of the
term $f_{NL}^{eq}$ during the first horizon crossing, to see what
the constant-roll condition brings along. The evaluation shall be
performed by using the values of the free parameters in such a way
so that the viability of the observational indices of inflation
is achieved according to the 2018 Planck data \cite{Akrami:2018odb}.

\section{TESTING THE THEORETICAL FRAMEWORK WITH THE LATEST OBSERVATIONAL DATA}

In this section we shall present explicitly examples of GW170817 compatible Einstein-Gauss-Bonnet  models, with the $f(R)$ extension, that can yield a phenomenologically viable inflationary era. First of all, the Gauss-Bonnet coupling scalar function $\xi(\phi)$ must be defined properly, aiming for a simple ratio $\xi'/\xi''$ so as to facilitate our study.

\subsection{Model compatible with the latest Planck Data under the slow-roll assumption}

Consider the coupling scalar
function $\xi(\phi)$ being equal to,
\begin{equation}
\centering
\xi(\phi)=\lambda_1 \int^{\kappa\phi} e^{\gamma_1(x)^n}dx,
\end{equation}
where $\lambda_1,\gamma_1$ and n are dimensionless constants to be specified later while $x$ is an auxiliary integration variable. This particular coupling scalar function is chosen due to the simplicity of the ratio $\xi' / \xi''$, specifically
\begin{equation}
\frac{\xi'}{\xi''}=\frac{ (\kappa \phi)^{1-n}}{\kappa n\gamma_1},
\end{equation}
even though the coupling itself seems superfluous. Nevertheless, it has been proven that it is a viable candidate for a Gauss-Bonnet scalar coupling function for the minimally coupled case so it is interesting to examine the impact of logarithmic corrections. One can specify the scalar potential from the equation (\ref{motion3final}), which has the following form,
\begin{equation}
\centering
V(\phi)=V_1 e^{\frac{(\kappa\phi)^{2-n}}{(2-n)n\gamma_1(\delta-1)}},
\end{equation}
where $V_1$ is the integration constant. Let us now proceed with the evaluation of the slow-roll indices,
\begin{equation}
\centering
\label{index1}
\epsilon_1\simeq\frac{(\kappa\phi)^{2-2n}}{2n^2\gamma_1^2},\,
\end{equation}
\begin{equation}
\label{index2}
\epsilon_2\simeq \frac{(\kappa \phi )^{-n} \left(\phi  V'-2 (n-1) V \right)}{2  \gamma_1 n V},\,
\end{equation}
\begin{equation}
\label{index3}
\centering
\epsilon_3\simeq -\frac{\delta  (\kappa \phi )^{2-2 n}}{2 \gamma_1^2 n^2 \left(\delta  \ln \left(-\frac{4 a \kappa^2 V}{\delta -1}\right)+\delta +1\right)} ,\,
\end{equation}
\begin{equation}
\label{index5}
\centering
\epsilon_5\simeq \frac{ (\kappa \phi )^{1-n} \left(8 \gamma_1  \lambda_1 n \kappa^4V (\kappa  \phi )^n e^{ \gamma_1 (\kappa \phi )^n}-3 (\delta -1) \delta  \kappa\phi \right)}{2\gamma_1   n \left(3 \gamma_1  (\delta -1) n (\kappa \phi )^n \left(\delta  \ln \left(-\frac{4 a \kappa^2 V}{\delta -1}\right)+\delta +1\right)+8  \lambda_1 \kappa\phi  \kappa^4V e^{\gamma_1  (\kappa \phi )^n}\right)} , \,
\end{equation}
\begin{equation}
\label{index6}
\centering
\epsilon_6\simeq\frac{  (\kappa \phi )^{1-n} \left(\kappa \phi \kappa^4 V \left(8 \lambda_1 \kappa^3V' e^{\gamma_1  (\kappa \phi )^n}-3 (\delta -1) \delta \right)+8 \gamma_1  (\kappa^4V)^2 e^{\gamma_1 (\kappa \phi )^n} \left(n \left(\gamma_1  (\kappa \phi )^n-1\right)+1\right)+3 \gamma_1  (\delta -1) \delta  n (\kappa \phi )^n \kappa^3V' \right)}{2 \gamma_1  n \kappa^4V \left(3 \gamma_1  (\delta -1) n (\kappa \phi )^n \left(\delta  \ln \left(-\frac{4 a \kappa^2 V}{\delta -1}\right)+\delta +1\right)+8 \lambda_1 \kappa\phi \kappa^4 V e^{\gamma_1  (\kappa \phi )^n}\right)} .\,
\end{equation}
\newline
It is obvious that the first three slow-roll indices have quite simple expressions in contrast to the next three due to the derivatives of the scalar potential and the derivatives of the coupling scalar function. Again $\epsilon_4$ is omitted due to the length expression. Now, we can determine the final value of the field when the inflationary era ends, by setting the the first slow-roll index Eq. (56) into unity,
\begin{equation}
\phi_f=\frac{ (2n^2 \gamma_1^2)^{\frac{1}{2-2n}}}{\kappa}.
\end{equation}
The initial value of the scalar field can be calculated from the e-folds number see, equation (\ref{efolds1})
\begin{equation}
\phi_i=\frac{((\kappa\phi_f)^2-\frac{N}{\gamma_1})^{\frac{1}{n}}}{\kappa},
\end{equation}
where we assumed that the number of e-folds is $N\simeq 60$. Considering the following values for the free parameters in Natural Units, namely, $\kappa^2=1$, $(N, n, \gamma_1, V_1,  \lambda_1, \delta,\alpha)=(60, -2, -5, 1, 1, 0.001, 1)$ then, the spectral index of primordial curvature perturbations $n_{\mathcal{S}}$, the tensor-to-scalar-ratio r and finally, the tensor spectral index $n_{\mathcal{T}}$ are respectively $n_{\mathcal{S}}=0.966381$, $r=0.000014$ and $n_{\mathcal{T}}=-1.7 \times 10^{-6}$, which are acceptable according to the latest Planck Data. Moreover, we found that the the initial value of the field is $\phi_i=0.236704$ and the final numerical value of the field is $\phi_f=2.41827$ which means that with the passage of time, based on the continuity, the field increases until the inflationary era ends. In addition, the numerical values of the slow-roll indices are $\epsilon_1=8.79 \times 10^{-7}$, $\epsilon_2=0.016$,   $\epsilon_3=\epsilon_4=\epsilon_5=-8.7 \times 10^{-10}$,   and $\epsilon_6=-1.75 \times 10^{-9}$, where it becomes apparent that all of them are subleading compared to the second slow-roll index.

\begin{figure}[h!]
\centering
\includegraphics[width=17pc]{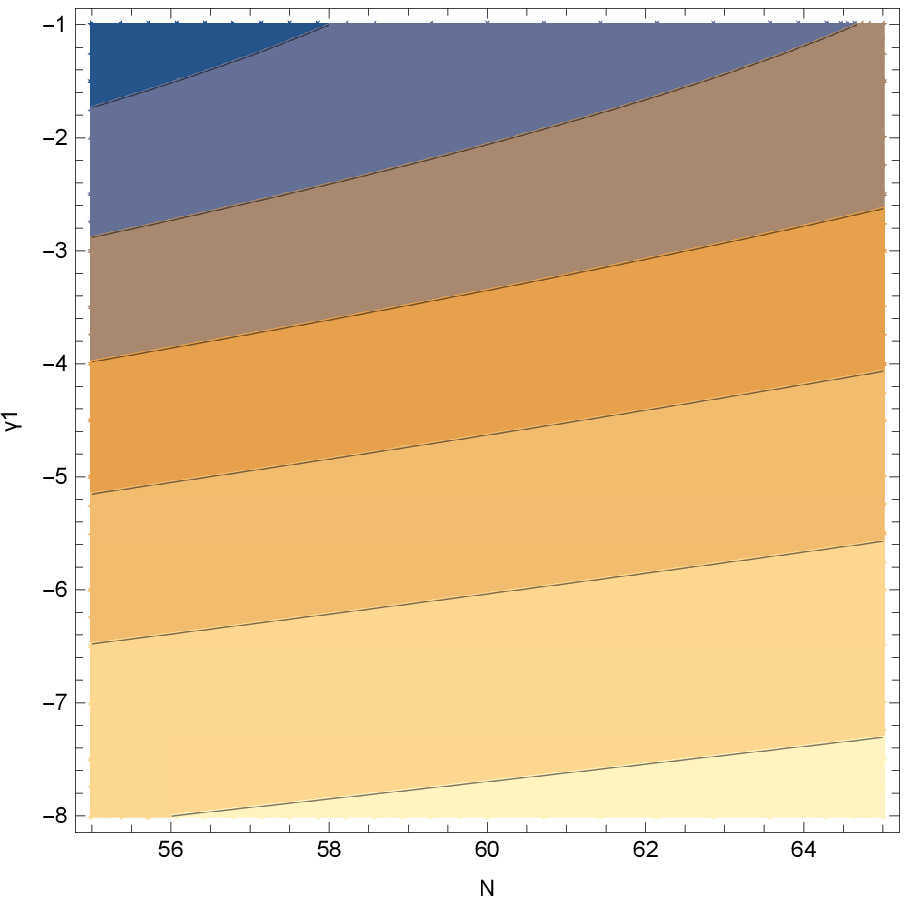}
\includegraphics[width=3pc]{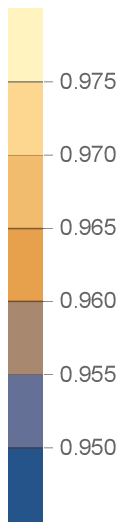}
\includegraphics[width=17pc]{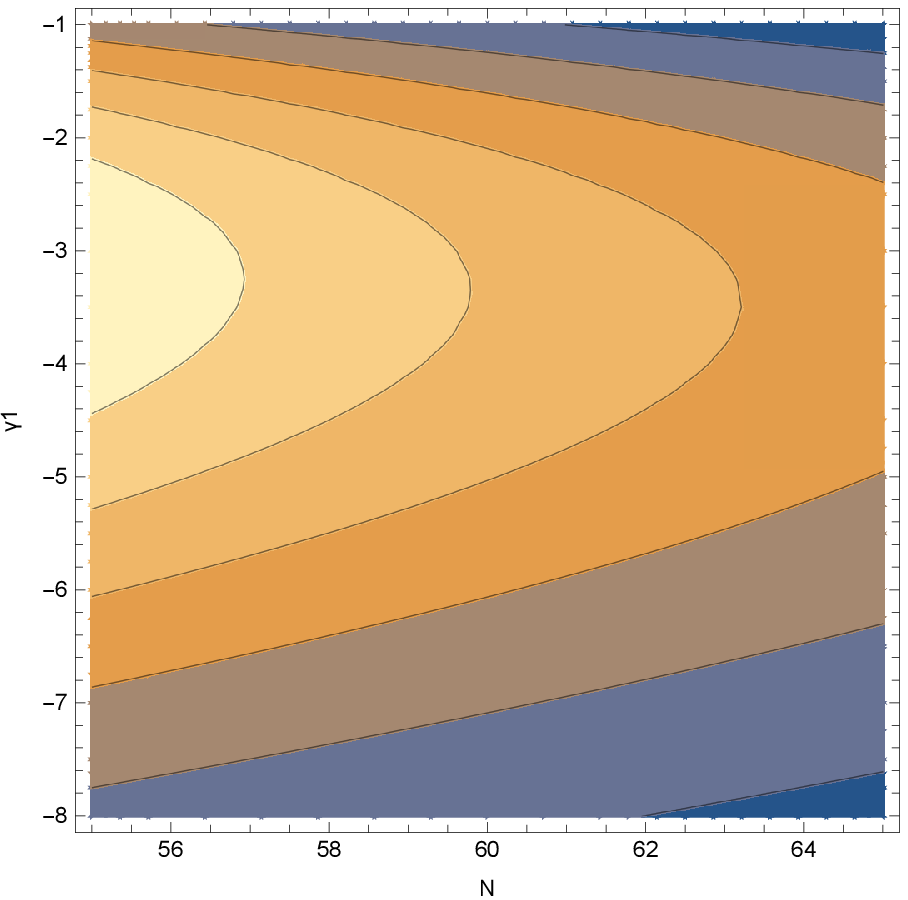}
\includegraphics[width=3pc]{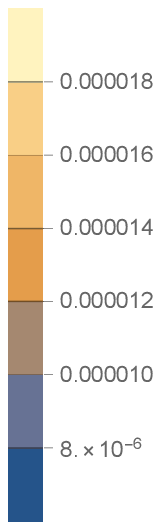}
\caption{Contour plots of the spectral index of primordial
curvature perturbations (left) and the tensor-to-scalar ratio
(right) depending on parameters $N$ and $\gamma_1$ ranging from
[55,65] and [-8,-1] respectively. It can be inferred that both
parameters influence their values but the spectral index changes
with a lesser rate.} \label{plot1}
\end{figure}

Lastly, we examine the validity of our approximations. Based on 
the slow-roll approximations, we note that $\dot
H\sim\mathcal{O}(10^{-7})$ compared to
$H^2\sim\mathcal{O}(10^{-1})$, similarly
$\frac{1}{2}\dot\phi^2\sim\mathcal{O}(10^{-7})$ in contrast
to $V\sim 1$. Indeed, the approximations in
(\ref{approx}) are valid. Moreover, the string terms in the
equations of motion (13) and (14) are
$24\dot\xi H^3\sim\mathcal{O}(10^{-42})$ and $16\dot\xi H\dot
H\sim\mathcal{O}(10^{-48})$ which justifies the reason the were
neglected. Furthermore, from the first modified Friedmann equation numerically speaking, $\frac{FR-f}{2\kappa^2}\sim \mathcal{O}(10^{-3})$ and $\frac{3H\dot F}{\kappa^2}\sim \mathcal{O}(10^{-9})$ are quite smaller than the scalar potential justifying our assumption. The term $\frac{\ddot F-H\dot F}{\kappa^2}\sim \mathcal{O}(10^{-10})$ in the second Friedmann equation is also minor in contrast to the term $\frac{2\dot HF}{\kappa^2}\sim \mathcal{O}(10^{-7})$.

\subsection{ Model with power-law coupling under the constant-roll approximation}

In this subsection it is considered a much simpler Gauss-Bonnet coupling scalar function $\xi(\phi)$ that follows a power-law form,
\begin{equation}
\centering
\label{coupling2}
\xi(\phi)=\lambda_2(\kappa \phi)^m.
\end{equation}
where $\lambda_2$ and $m$ are dimensionless constants to be specified later. This is a very appealing function since the ratio $\frac{\xi'}{\xi''}$  which appears in our calculations is greatly simplified, since,
\begin{equation}
\frac{\xi'}{\xi''}=\frac{\phi}{m-1}.
\end{equation}
This model was also studied in, Ref. \cite{Odintsov:2020mkz}. Since the scalar coupling function is specified, one can find the scalar potential from Eq. (\ref{motion3}),
\begin{equation}
\centering
V(\phi)=V_2e^{-\frac{(\beta^2+2\beta-3)(\kappa\phi)^2}{6(m-1)(\delta-1)}},
\end{equation}
where $V_2$ is an integration constant. For this model the constant-roll indices are given by the following equations,
\begin{equation}
\centering
\label{index1}
\epsilon_1=\frac{(\beta-1)^2 (\kappa\phi) ^2}{2 (m-1)^2} ,\,
\end{equation}
\begin{equation}
\label{index2}
\epsilon_2=\beta ,\,
\end{equation}
\begin{equation}
\label{index3}
\centering
\epsilon_3=-\frac{(\beta-1)^2 \delta (\kappa\phi) ^2}{2 (m-1)^2 \left(\delta  \ln \left(-\frac{4 a \kappa^2 V(\phi )}{\delta -1}\right)+\delta +1\right)} ,\,
\end{equation}
\begin{equation}
\label{index5}
\centering
\epsilon_5=  \frac{(\beta-1) \left(3 (\beta-1) (\delta -1) \delta  (\kappa\phi) ^2+8 \lambda_2 (m-1) m \kappa^4V (\kappa \phi )^m\right)}{2 (m-1) \left(8 (\beta-1) \lambda_2 m \kappa^4V (\kappa \phi )^m-3 (\delta -1) (m-1) \left(\delta  \ln \left(-\frac{4 a \kappa^2 V}{\delta -1}\right)+\delta +1\right)\right)}, \,
\end{equation}
\begin{equation}
\label{index6}
\centering
\epsilon_6=-\frac{(\beta-1) \left(8 (\beta-1) \lambda_2 m \kappa^4V (\kappa \phi )^m \left(m \kappa^4V+\kappa\phi  \kappa^3V' \right)-3 (\delta -1) \delta  \phi  \left((\beta-1) \kappa \phi \kappa^4 V +(m-1) \kappa^3V'\right)\right)}{2 (m-1) \kappa^4V \left(8 (\beta-1)  \lambda_2 m \kappa^4V (\kappa \phi )^m-3 (\delta -1) (m-1) \left(\delta  \ln \left(-\frac{4 a \kappa^2 V}{\delta -1}\right)+\delta +1\right)\right)}, \,
\end{equation}
where $\epsilon_4$ is not written analytically due to the perplexed form. Despite the fact that, the indices $\epsilon_4$ to $\epsilon_6$ are quite perplexed, the indices $\epsilon_1$ to $\epsilon_3$ have very simple forms. By setting the index $\epsilon_1$ to unity, the final value of the field when inflationary era ends is equal to,
\begin{equation}
\phi_f=\pm \frac{\sqrt{2}}{\kappa\sqrt{\frac{(\beta-1)^2}{(m-1)^2}}},
\end{equation}
while, the initial value of the field is,
\begin{equation}
\phi_i=\pm \phi_f e^{-\frac{N(1-\beta)}{m-1}}.
\end{equation}
Our following analysis is based in the case of the positive value of the scalar field.  Assigning the following values
to the free parameters, always in reduced Planck units, $(\kappa, \delta, \lambda_2, m, N, \beta, V_2, \alpha)=(1, 0.003, 1, 8, 60,  0.017, 1, 1)$ then, the observational indices take the following values  $n_{\mathcal{S}}=0.966$, $r=7.6 \times 10^{-7}$ and $n_{\mathcal{T}}=-9.5 \times 10^{-8} $, which are acceptable according to the latest Planck Data. Furthermore $c_A=1$, hence, the model is free of ghosts. The numerical values of the slow-roll indices are $\epsilon_1\simeq 4.8 \times 10^{-8}$, $\epsilon_2=0.017$, $\epsilon_3=-2.3 \times 10^{-10}$ and the rest indices are all equal to $\epsilon_3$, a feature which is expected at least for $\epsilon_5$ and $\epsilon_6$ when $F$ is more dominant than string corrections. Moreover, we mention that the initial and final value of the scalar field are $\phi_i\simeq 0.002$
and $\phi_f=10.0707$ which indicates an increase in the scalar field. In Fig. 2 we plot the spectral index of primordial
curvature perturbations $n_\mathcal{S}$ (left) and the tensor-to-scalar ratio r (right) depending on parameters $\beta$ and m ranging [0.001, 0.009] and [8, 12] respectively.

\begin{figure}[h!]
\centering
\includegraphics[width=17pc]{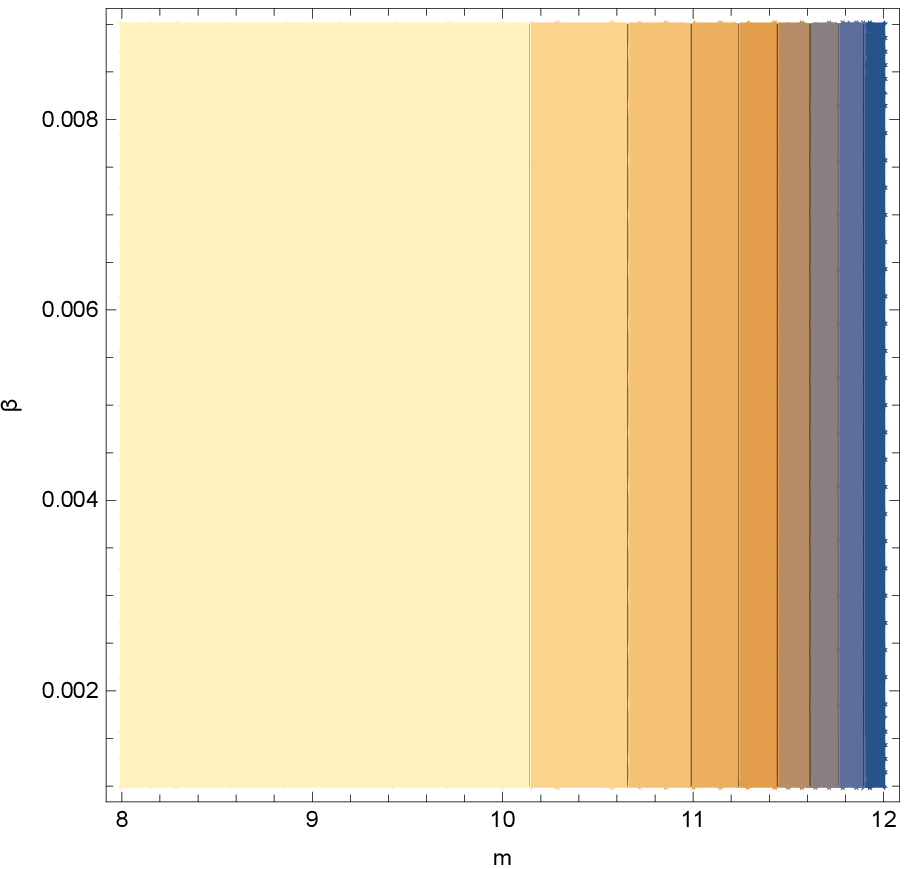}
\includegraphics[width=3pc]{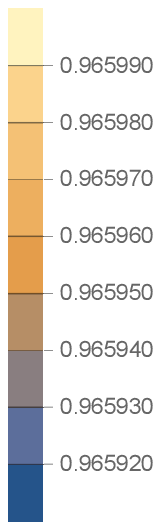}
\includegraphics[width=17pc]{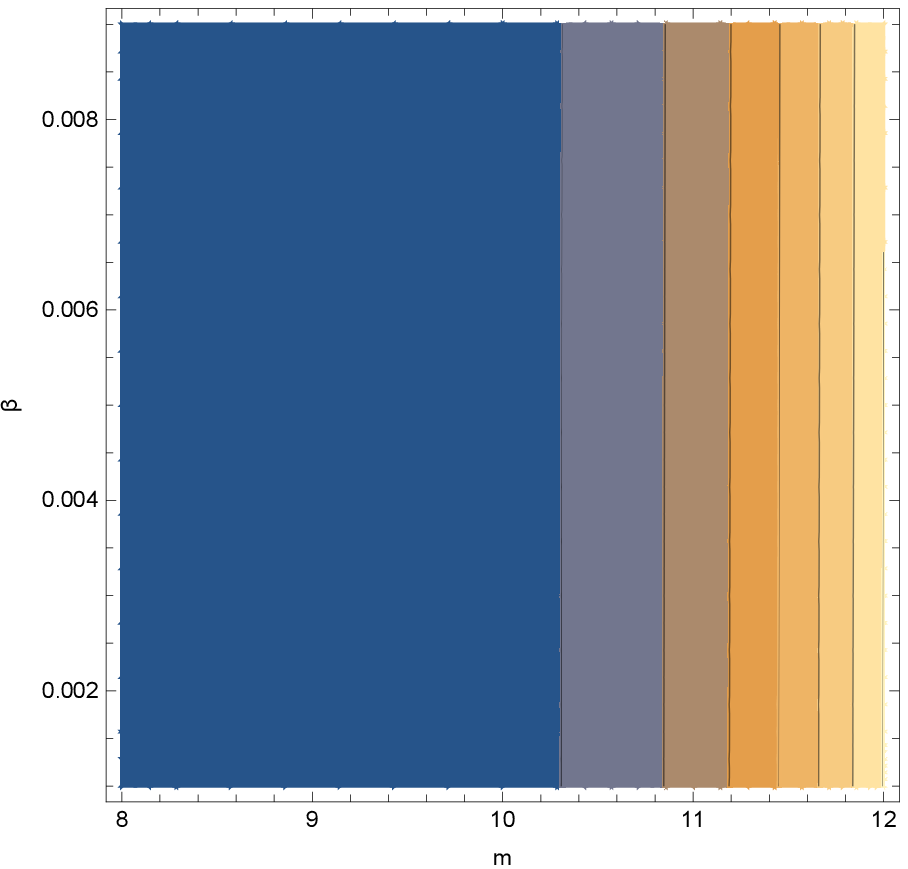}
\includegraphics[width=3pc]{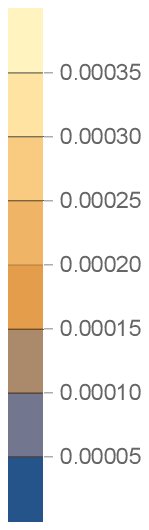}
\caption{Contour plots of the spectral index of primordial
curvature perturbations (left) and the tensor-to-scalar ratio
(right) depending on parameters $\beta$ and $m$ ranging from
[0.001, 0.009] and [8, 12] respectively. Concerning the spectral index, it is clear
that the dominant parameter which defines its value is the constant-roll parameter.}
\end{figure}

Moreover, we make predictions for the amount of
non-Gaussianities in the primordial power spectrum of the
curvature perturbations. The auxiliary terms of the specific model are defined as,
\begin{equation}
\centering
\delta_\xi\simeq \frac{(\beta-1) \lambda_2 m \kappa^4V (\kappa \phi )^m}{3 (\delta -1) (m-1)}\,
\end{equation}
\begin{equation}
\centering
\epsilon_s\simeq\frac{(\beta-1) \left(3 (\beta-1) (\delta -1) (\kappa\phi) ^2-8  \lambda_2 (m-1) m \kappa^4V (\kappa \phi )^m\right)}{6 (\delta -1) (m-1)^2} ,\,
\end{equation}
\begin{equation}
\centering
n_s\simeq \frac{8 (\beta-1)  \lambda_2 (m-1) m (\kappa \phi )^m \left(m \kappa^4V+\kappa\phi \kappa^3 V' \right)-6 (\beta-1)^2 (\delta -1) (\kappa\phi) ^2}{(m-1) \left(3 (\beta-1) (\delta -1) (\kappa\phi) ^2-8 \lambda_2 (m-1) m \kappa^4V (\kappa \phi )^m\right)} .\,
\end{equation}
From equation (\ref{NL}), the expected
value of $f_{NL}^{eq}$, for the exact same set of parameters we
used to obtain the viability of the model with the Planck data, is
$f_{NL}^{eq}=0.117024$ which is also an accepted value and may
explain why non Gaussianities have yet to be observed. Finally,
the parameters used to derive such values are 
$\delta_\xi\simeq 2.1 \times 10^{-22}$, $\epsilon_s=4.80305 \times 10^{-8}$ and
$\eta=0.280857$ which means that one of them is in fact
dominant. These results imply that $\epsilon_s=\epsilon_1$.

At this point, it is important to mention that, the observed quantities $n_S$ and $r$  experience different changes after the alteration of the free parameters. Specifically, the primordial spectral index is affected only by the   constant-roll parameter $\beta$ while,  the tensor-to-scalar ratio, is affected by the exponent $m$ of the coupling scalar function along with the constant-roll parameter with the first being
more dominant factor. This can easily be observed in Fig.
2 where one sees that the spectral index of scalar
perturbations is depicted by a simple plot resembling vertical
lines. In addition, while the term $f_{NL}^{eq}$ is independent of
parameter $\lambda_2$, it can be enhanced by
decreasing the exponent $m$ but such a change leads to a subsequent decrease in the
tensor-to-scalar ratio. For instance, numerically speaking, choosing $m=2$ leads to
$f_{NL}^{eq}=0.819167$, $n_S=0.966$ and the effective value of the
tensor-to-scalar ratio is 0, since, $r\sim\mathcal{O}(10^{-51})$. 

Finally, we examine the validity of the approximations which were
made during this model at the first horizon crossing. Firstly, the
slow-roll approximations in Eq. (\ref{approx}) hold true since
$\dot H\sim\mathcal{O}(10^{-8})$ compared to
$H^2\sim\mathcal{O}(10^{-1})$ and
$\frac{1}{2}\dot\phi^2\sim\mathcal{O}(10^{-8})$ compared to
$V\sim\mathcal{O}(10^{1})$ are negligible. Also, the terms which
were omitted in equations (\ref{motion1}) and (\ref{motion2}) are
of of the order (in reduced Planck units) $24\dot\xi
H^3\sim\mathcal{O}(10^{-21})$ while $16\dot\xi H\dot
H\sim\mathcal{O}(10^{-29})$, which explains why these terms,
compared to the scalar potential and the kinetic term, can be
neglected and $V'\sim\mathcal{O}(10^{-4})$ whereas
$\xi'\mathcal{G}\sim\mathcal{O}(10^{-18})$ which explains the neglect  of string term. Lastly, the following two terms in equation (\ref{motion1}) $\frac{FR-f}{2\kappa^2}\sim \mathcal{O}(10^{-3})$ and $\frac{3H\dot F}{k^2}\sim \mathcal{O}(10^{-10})$ are quite smaller than the scalar potential and the term $\frac{\ddot F-H\dot F}{k^2}\sim \mathcal{O}(10^{-11})$ in the second Friedmann equation is also minor in contrast to the term $\frac{2\dot HF}{\kappa^2}\sim \mathcal{O}(10^{-8})$ justifying our approximations.

\section{Conclusions}

In this work it is presented an alteration of our previous work \cite{Odintsov:2020sqy}, by using a modified $f(R)$ logarithmic gravity instead of Einstein's gravity, consistent with the GW170817 event. We focused our analysis on the inflationary era of the universe, by considering that the scalar field evolves with either under slow-roll assumptions or with a constant rate of roll. In both of these cases, the slow-roll indices and the observational quantities of inflation were evaluated along with the predicted amount of Non-Gaussianities, in the case of  constant-roll evolution of the scalar field. After the theoretical framework we confronted two models with the observational data coming from the Planck 2018 collaboration, considering logarithmic corrections.
As shown, the resulting inflationary phenomenology can
be compatible with the latest Planck data, for a wide range of the free parameters of the theory, after neglecting string corrections, which is shown to be a reasonable approach. In our analysis we demonstrated that all the assumptions made were satisfied for
all the models examined, and for the values of the free parameters that yield inflationary viability with respect to
the latest Planck data. In the constant-roll case we also
investigated the amount of non-Gaussianities that are predicted
from the model, by calculating the nonlinear term $f_{NL}^{eq}$ in the
equilateral momentum approximation. Interestingly enough, we
demonstrated that the amount of non-Gaussianities is quite small.
Finally, we
performed an analytic approximation in the differential equation
that connects the scalar field potential and the scalar coupling
function, and we examined the phenomenology of inflation in this
case too. As evinced, the models can also be compatible with the
Planck 2018 too, even with a logarithmic modified gravity $f(R)$ with an almost linear Ricci scalar through logarithmic corrections.


\begin{thebibliography}{99}



\bibitem{GBM:2017lvd}
  B.~P.~Abbott {\it et al.}
  ``Multi-messenger Observations of a Binary Neutron Star Merger,''
  Astrophys.\ J.\  {\bf 848} (2017) no.2,  L12
  doi:10.3847/2041-8213/aa91c9
  [arXiv:1710.05833 [astro-ph.HE]].




\bibitem{Ezquiaga:2017ekz}
  J.~M.~Ezquiaga and M.~Zumalacarregui,
  Phys.\ Rev.\ Lett.\  {\bf 119} (2017) no.25,  251304
  doi:10.1103/PhysRevLett.119.251304
  [arXiv:1710.05901 [astro-ph.CO]].










\bibitem{Hwang:2005hb}
  J.~c.~Hwang and H.~Noh,
  Phys.\ Rev.\ D {\bf 71} (2005) 063536
  doi:10.1103/PhysRevD.71.063536
  [gr-qc/0412126].


\bibitem{Nojiri:2006je}
  S.~Nojiri, S.~D.~Odintsov and M.~Sami,
  Phys.\ Rev.\ D {\bf 74} (2006) 046004
  doi:10.1103/PhysRevD.74.046004
  [hep-th/0605039].




\bibitem{Cognola:2006sp}
  G.~Cognola, E.~Elizalde, S.~Nojiri, S.~Odintsov and S.~Zerbini,
  Phys.\ Rev.\ D {\bf 75} (2007) 086002
  doi:10.1103/PhysRevD.75.086002
  [hep-th/0611198].



\bibitem{Nojiri:2005vv}
  S.~Nojiri, S.~D.~Odintsov and M.~Sasaki,
  Phys.\ Rev.\ D {\bf 71} (2005) 123509
  doi:10.1103/PhysRevD.71.123509
  [hep-th/0504052].


\bibitem{Nojiri:2005jg}
  S.~Nojiri and S.~D.~Odintsov,
  Phys.\ Lett.\ B {\bf 631} (2005) 1
  doi:10.1016/j.physletb.2005.10.010
  [hep-th/0508049].







\bibitem{Satoh:2007gn}
  M.~Satoh, S.~Kanno and J.~Soda,
  Phys.\ Rev.\ D {\bf 77} (2008) 023526
  doi:10.1103/PhysRevD.77.023526
  [arXiv:0706.3585 [astro-ph]].



\bibitem{Bamba:2014zoa}
  K.~Bamba, A.~N.~Makarenko, A.~N.~Myagky and S.~D.~Odintsov,
  JCAP {\bf 1504} (2015) 001
  doi:10.1088/1475-7516/2015/04/001
  [arXiv:1411.3852 [hep-th]].


\bibitem{Yi:2018gse}
  Z.~Yi, Y.~Gong and M.~Sabir,
  Phys.\ Rev.\ D {\bf 98} (2018) no.8,  083521
  doi:10.1103/PhysRevD.98.083521
  [arXiv:1804.09116 [gr-qc]].


\bibitem{Guo:2009uk}
  Z.~K.~Guo and D.~J.~Schwarz,
  Phys.\ Rev.\ D {\bf 80} (2009) 063523
  doi:10.1103/PhysRevD.80.063523
  [arXiv:0907.0427 [hep-th]].


\bibitem{Guo:2010jr}
  Z.~K.~Guo and D.~J.~Schwarz,
  Phys.\ Rev.\ D {\bf 81} (2010) 123520
  doi:10.1103/PhysRevD.81.123520
  [arXiv:1001.1897 [hep-th]].


\bibitem{Jiang:2013gza}
  P.~X.~Jiang, J.~W.~Hu and Z.~K.~Guo,
  Phys.\ Rev.\ D {\bf 88} (2013) 123508
  doi:10.1103/PhysRevD.88.123508
  [arXiv:1310.5579 [hep-th]].



\bibitem{Kanti:2015pda}
  P.~Kanti, R.~Gannouji and N.~Dadhich,
  Phys.\ Rev.\ D {\bf 92} (2015) no.4,  041302
  doi:10.1103/PhysRevD.92.041302
  [arXiv:1503.01579 [hep-th]].



\bibitem{Nojiri:2017ncd}
S.~Nojiri, S.~D.~Odintsov and V.~K.~Oikonomou,
Phys. Rept. \textbf{692} (2017), 1-104
doi:10.1016/j.physrep.2017.06.001
[arXiv:1705.11098 [gr-qc]].


\bibitem{Kanti:1998jd}
  P.~Kanti, J.~Rizos and K.~Tamvakis,
  Phys.\ Rev.\ D {\bf 59} (1999) 083512
  doi:10.1103/PhysRevD.59.083512
  [gr-qc/9806085].




\bibitem{Pozdeeva:2020apf}
  E.~O.~Pozdeeva, M.~R.~Gangopadhyay, M.~Sami, A.~V.~Toporensky and S.~Y.~Vernov,
  arXiv:2006.08027 [gr-qc].

\bibitem{Fomin:2020hfh}
  I.~Fomin,
  arXiv:2004.08065 [gr-qc].

\bibitem{DeLaurentis:2015fea}
  M.~De Laurentis, M.~Paolella and S.~Capozziello,
  Phys.\ Rev.\ D {\bf 91} (2015) no.8,  083531
  doi:10.1103/PhysRevD.91.083531
  [arXiv:1503.04659 [gr-qc]].


\bibitem{Chervon:2019sey}
  S.~Chervon, I.~Fomin, V.~Yurov and A.~Yurov,
  doi:10.1142/11405



\bibitem{Nozari:2017rta}
  K.~Nozari and N.~Rashidi,
  Phys.\ Rev.\ D {\bf 95} (2017) no.12,  123518
  doi:10.1103/PhysRevD.95.123518
  [arXiv:1705.02617 [astro-ph.CO]].




\bibitem{Odintsov:2018zhw}
  S.~D.~Odintsov and V.~K.~Oikonomou,
  Phys.\ Rev.\ D {\bf 98} (2018) no.4,  044039
  doi:10.1103/PhysRevD.98.044039
  [arXiv:1808.05045 [gr-qc]].


  \bibitem{Kawai:1998ab}
  S.~Kawai, M.~a.~Sakagami and J.~Soda,
  Phys.\ Lett.\ B {\bf 437}, 284 (1998)
  doi:10.1016/S0370-2693(98)00925-3
  [gr-qc/9802033].


\bibitem{Yi:2018dhl}
  Z.~Yi and Y.~Gong,
  Universe {\bf 5} (2019) no.9,  200
  doi:10.3390/universe5090200
  [arXiv:1811.01625 [gr-qc]].


\bibitem{vandeBruck:2016xvt}
  C.~van de Bruck, K.~Dimopoulos and C.~Longden,
  Phys.\ Rev.\ D {\bf 94} (2016) no.2,  023506
  doi:10.1103/PhysRevD.94.023506
  [arXiv:1605.06350 [astro-ph.CO]].


\bibitem{Kleihaus:2019rbg}
  B.~Kleihaus, J.~Kunz and P.~Kanti,
  arXiv:1910.02121 [gr-qc].





\bibitem{Bakopoulos:2019tvc}
  A.~Bakopoulos, P.~Kanti and N.~Pappas,
  Phys.\ Rev.\ D {\bf 101} (2020) no.4,  044026
  doi:10.1103/PhysRevD.101.044026
  [arXiv:1910.14637 [hep-th]].


\bibitem{Maeda:2011zn}
  K.~i.~Maeda, N.~Ohta and R.~Wakebe,
  Eur.\ Phys.\ J.\ C {\bf 72} (2012) 1949
  doi:10.1140/epjc/s10052-012-1949-6
  [arXiv:1111.3251 [hep-th]].






\bibitem{Bakopoulos:2020dfg}
  A.~Bakopoulos, P.~Kanti and N.~Pappas,
  arXiv:2003.02473 [hep-th].


\bibitem{Ai:2020peo}
W.~Ai,
[arXiv:2004.02858 [gr-qc]].



\bibitem{Odintsov:2019clh}
  S.~D.~Odintsov and V.~K.~Oikonomou,
  Phys.\ Lett.\ B {\bf 797} (2019) 134874
  doi:10.1016/j.physletb.2019.134874
  [arXiv:1908.07555 [gr-qc]].



\bibitem{Oikonomou:2020oil}
V.~K.~Oikonomou and F.~P.~Fronimos,
[arXiv:2007.11915 [gr-qc]].

\bibitem{Odintsov:2020xji}
S.~D.~Odintsov, V.~K.~Oikonomou and F.~P.~Fronimos,
Annals Phys. \textbf{420} (2020), 168250
doi:10.1016/j.aop.2020.168250 [arXiv:2007.02309 [gr-qc]].



\bibitem{Oikonomou:2020sij}
V.~K.~Oikonomou and F.~P.~Fronimos,
[arXiv:2006.05512 [gr-qc]].



\bibitem{Odintsov:2020zkl}
S.~D.~Odintsov and V.~K.~Oikonomou,
Phys. Lett. B \textbf{805} (2020), 135437
doi:10.1016/j.physletb.2020.135437 [arXiv:2004.00479 [gr-qc]].


\bibitem{Odintsov:2020sqy}
S.~D.~Odintsov, V.~K.~Oikonomou and F.~P.~Fronimos,
[arXiv:2003.13724 [gr-qc]].





\bibitem{Easther:1996yd}
  R.~Easther and K.~i.~Maeda,
  Phys.\ Rev.\ D {\bf 54} (1996) 7252
  doi:10.1103/PhysRevD.54.7252
  [hep-th/9605173].

\bibitem{Antoniadis:1993jc}
  I.~Antoniadis, J.~Rizos and K.~Tamvakis,
  Nucl.\ Phys.\ B {\bf 415} (1994) 497
  doi:10.1016/0550-3213(94)90120-1
  [hep-th/9305025].

\bibitem{Antoniadis:1990uu}
I.~Antoniadis, C.~Bachas, J.~R.~Ellis and D.~V.~Nanopoulos,
Phys.\ Lett.\ B \textbf{257} (1991), 278-284
doi:10.1016/0370-2693(91)91893-Z




\bibitem{Kanti:1995vq}
P.~Kanti, N.~Mavromatos, J.~Rizos, K.~Tamvakis and E.~Winstanley,
Phys. Rev. D \textbf{54} (1996), 5049-5058
doi:10.1103/PhysRevD.54.5049 [arXiv:hep-th/9511071 [hep-th]].



\bibitem{Kanti:1997br}
P.~Kanti, N.~Mavromatos, J.~Rizos, K.~Tamvakis and E.~Winstanley,
Phys. Rev. D \textbf{57} (1998), 6255-6264
doi:10.1103/PhysRevD.57.6255 [arXiv:hep-th/9703192 [hep-th]].


\bibitem{Bajardi:2019zzs}
  F.~Bajardi, K.~F.~Dialektopoulos and S.~Capozziello,
  Symmetry {\bf 12} (2020) no.3,  372
  doi:10.3390/sym12030372
  [arXiv:1911.03554 [gr-qc]].


\bibitem{Capozziello:2019wfi}
  S.~Capozziello, C.~A.~Mantica and L.~G.~Molinari,
  Int.\ J.\ Geom.\ Meth.\ Mod.\ Phys.\  {\bf 16} (2019) no.09,  1950133
  doi:10.1142/S0219887819501330
  [arXiv:1906.05693 [gr-qc]].


\bibitem{deMartino:2020yhq}
I.~de Martino, M.~De Laurentis and S.~Capozziello,
[arXiv:2008.09856 [gr-qc]].


\bibitem{Nojiri:2010wj}
  S.~Nojiri and S.~D.~Odintsov,
  Phys.\ Rept.\  {\bf 505} (2011) 59
  doi:10.1016/j.physrep.2011.04.001
  [arXiv:1011.0544 [gr-qc]].
  
  
\bibitem{Odintsov:2020mkz}
S.~D.~Odintsov, V.~K.~Oikonomou, F.~P.~Fronimos and S.~A.~Venikoudis,
Phys. Dark Univ. \textbf{30} (2020), 100718
doi:10.1016/j.dark.2020.100718
[arXiv:2009.06113 [gr-qc]].
  
\bibitem{Akrami:2018odb}
Y.~Akrami \textit{et al.} [Planck],
Astron. Astrophys. \textbf{641} (2020), A10
doi:10.1051/0004-6361/201833887
[arXiv:1807.06211 [astro-ph.CO]].
  
\bibitem{Odintsov:2017hbk}
S.~D.~Odintsov, V.~K.~Oikonomou and L.~Sebastiani,
Nucl. Phys. B \textbf{923} (2017), 608-632
doi:10.1016/j.nuclphysb.2017.08.018
[arXiv:1708.08346 [gr-qc]].
  
\bibitem{Nojiri:2003ni}
S.~Nojiri and S.~D.~Odintsov,
Gen. Rel. Grav. \textbf{36} (2004), 1765-1780
doi:10.1023/B:GERG.0000035950.40718.48
[arXiv:hep-th/0308176 [hep-th]].
  
\bibitem{DeFelice:2011zh}
A.~De Felice and S.~Tsujikawa,
JCAP \textbf{04} (2011), 029
doi:10.1088/1475-7516/2011/04/029
[arXiv:1103.1172 [astro-ph.CO]].
  
  
  
  
\end{thebibliography}
\end{document}